\documentclass{elsart}
\usepackage{epsfig}

\begin{document}

\begin{frontmatter}
\title{Evidence for Power-law tail of the Wealth Distribution in India}

\author{Sitabhra Sinha}

\address{The Institute of Mathematical Sciences, C. I. T. Campus,
Taramani, Chennai - 600 113, India.}

\begin{abstract}
The higher-end tail of the wealth distribution in India is studied
using recently published lists of the wealth of richest Indians between 
the years 2002-4. The resulting rank distribution seems to imply a power-law
tail for the wealth distribution, with a Pareto exponent between 0.81
and 0.92 (depending on the year under analysis). This provides a
comparison with previous studies of wealth distribution, which have
all been confined to Western advanced capitalist economies. We
conclude with a discussion on the appropriateness of multiplicative
stochastic process as a model for asset accumulation, 
the relation between the wealth and income
distributions (we estimate the Pareto exponent for the latter to be
around 1.5 for India), 
as well as possible sources of error in measuring the
Pareto exponent for wealth.

\vspace{0.25cm}
\noindent
PACS numbers: 89.65.Gh, 89.65.-s, 02.50.-r, 89.75.Da
\end{abstract}

\end{frontmatter}

\vspace{-0.4cm}
\section{Introduction}

\vspace{-0.7cm}
More than a century ago, Pareto had observed that the income
distribution across several countries (at least in the high-income range)
follows a power law \cite{Par97}, i.e., the probability density 
function of income $I$,
$P( I ) \sim I^{- ( 1 + \alpha )}$, with the Pareto exponent $\alpha$
lying between 1 and 2. Pareto claimed that, in general, $\alpha \sim
1.5$. The power-law nature was also found to be true of wealth distributions, 
albeit
with a different exponent.
The two distributions are not completely unrelated, as those who 
are significantly wealthy also have incomes far
higher than the average individual or household.
However, the distributions of income and wealth
cannot be simply connected, and each have to be measured 
independently
for a particular society.
The occurrence of a qualitatively similar distribution across
widely differing geographical regions and economic development stages
may be indicative of universal features of inequality in human
societies.
This has led to attempts at developing simple models for generating
wealth distributions that are 
qualitatively similar to those empirically observed, 
with asset exchange
interactions between agents
\cite{Isp98,Dra00,Sin03,Cha04,Sla04,Sin05}. To verify such models
further empirical measurements of wealth distribution in different
economies is essential.

\vspace{-0.3cm}
Very recently, there have been a large number of empirical studies of 
the income distribution of several countries, with income being
defined as the flow of wages,
dividends, interest payments, etc. over a period of time. This can
usually be inferred from income tax returns. The general consensus,
based on these studies, is that at the low-income range the income
distribution obeys a 
log-normal \cite{Sou01} or exponential \cite{Dra01,Dra01a} distribution, 
while the high-income end shows power law
behavior with widely differing Pareto exponents, which are different
both in different countries, as well as in different periods for the
same country (e.g., see Ref. \cite{Fuj03}). 

\vspace{-0.3cm}
Unfortunately, 
not many studies
have been done on the distribution of wealth, which consist of the net
value of assets (financial holdings and/or tangible items) owned at 
a given point in time. 
The lack of an easily available data source for measuring wealth,
analogous to income tax returns for measuring income, means that one
has to resort to indirect methods. Levy and Solomon \cite{Lev97} used
a published list of wealthiest people to generate a rank-order distribution,
from which they inferred the Pareto exponent for wealth distribution
in USA. Follow-up studies used similar techniques to infer the
exponents for UK, France and Sweden \cite{Lev98,Lev03}. Refs. \cite{Dra01} and
\cite{Coe04} used an alternative technique based on adjusted data 
reported for the purpose of inheritance tax to obtain the Pareto 
exponent for UK.
Another study used tangible asset (namely house area) as a measure of
wealth to obtain the wealth distribution exponent in ancient Egyptian
society during the reign of Akhenaten (14th century BC)\cite{Abu02}.
Apart from the last mentioned study, all the other wealth distributions
were for western highly-developed capitalist economies, and
are thus of very similar societies. Observing the wealth distribution
of a non-Western developing capitalist society, such as India, 
which until quite 
recently had a planned economy, will be not only instructive by itself
but it will also provide necessary comparison with the previous studies.

\vspace{-0.3cm}
The general feature observed in the limited empirical study of 
wealth distribution is that of a power law behavior for the wealthiest
5-10 $\%$ of the population, and exponential or log-normal
distribution for the rest of the population.
The Pareto exponent as measured from the wealth distribution is found
to be always lower than the exponent for the income distribution,
which is consistent with the general observation that, in market
economies, wealth is much more unequally distributed than 
income \cite{Sam01}. 

\vspace{-0.3cm}
In the present paper, we have observed that the high wealth limit of the
Indian wealth distribution is consistent with a power law
having an exponent that ranges from 0.81 (2002) to 0.92
(2004). In the next section we describe the data sets used in our
analysis. In the section containing results we have reported not only
the power law behavior, but also how changes in wealth is related to
ones net worth. Data on labor income (salaries) at the top-income
end is also analyzed and compared with the low- and middle-income distribution.
We conclude with a discussion on the reliability of exponent
measurements, possible reasons for obtaining 
multiple values of the Pareto exponent for the same economy, and the
connection with such low-resolution measure of inequality as the Gini 
coefficient.

\vspace{-0.8cm}
\section{Data Sources}

\vspace{-0.7cm}
The data for the 125 wealthiest individuals and households in India were
obtained from a special report by the Indian business magazine, {\em
Business Standard}\cite{BusSta}. The wealths were reported at two dates, Dec
31, 2002 and Aug 31, 2003, which allowed us to also study the change
in wealth over the interval between these two dates. The
list essentially comprised of Indian billionaires (in Indian Rupees)
as of Aug 31, 2003. For comparison, note that India had 61,000 millionaires 
in 2003 \cite{MerLyn}; by contrast, USA had 2,270,000
millionaires.

\vspace{-0.3cm}
The above data set also reported the gross salary of the 67 highest-paid
executives in India (which includes foreign nationals based in
India). Many, though by no means all, of those who figure
in this list also belong to the previously mentioned list of
wealthiest Indians. It is therefore possible to infer a relation between
labor income and wealth.

\vspace{-0.3cm}
We also used a recent list of 40 richest Indians published by the
international business magazine {\em Forbes} in Dec 10, 2004 \cite{Forbes}.
The criterion used for this
list was somewhat different from the {\em Business Standard} list in
that an individual did not need to be residing in India to be listed, but need
only have Indian nationality. However, in practice, except for one
case, all the others in the list are based in India. Further, while in
the previous list the wealth was calculated in Indian Rupees, in the
Forbes list it is given in terms of US Dollars. However, as we are
primarily interested in the slope of the rank-order distribution, 
this did not affect our results.

\vspace{-0.3cm}
We rejected the top $10\%$ of the data in the lists while fitting a
power-law function to the distributions.
This was to avoid erroneous calculation of the exponent due to the
wealth of the richest few individuals being higher than the general
trend, resembling the `King effect' seen in many other contexts, e.g.,
the distribution of city sizes \cite{Sor98}, popularity of musicians 
\cite{Dav02}, movie gross earnings \cite{Sin04}, etc.
We also classified the wealths according to the industry sectors on
which they were founded. The classification was adopted from the
Bombay Stock Exchange (BSE) list of 20 industry sectors, ranging from
information technology, pharmaceuticals, automotive, etc. which have a
large representation in the list of wealthiest Indians, to sectors such
as, food \& beverages, consumer durables, consumer non-durables,
etc., each of which have so few representatives in the sample, that 13
of them have been grouped together into an aggregation called `Others' 
in our study.

\section{Results}
\vspace{-0.7cm}
As pointed out in previous papers (e.g., see Ref. \cite{Sin04}), the exponent
of a power-law probability distribution function can be determined with
good accuracy from the slope of the corresponding rank-order plot on a 
double logarithmic scale. In particular, if the wealth is distributed as
$P ( W ) \sim W^{- ( 1 + \alpha )}$, it can be shown that the wealth of the
$k$-th ranked agent is distributed as $W_k \sim k^{-1 / \alpha}$ \cite{Red98}.
Hence, obtaining the slope of the rank-order plot on a
double logarithmic scale and inverting it, allows us to determine the
Pareto exponent.
\begin{figure}
\epsfxsize=5.5in
\centerline{\epsffile{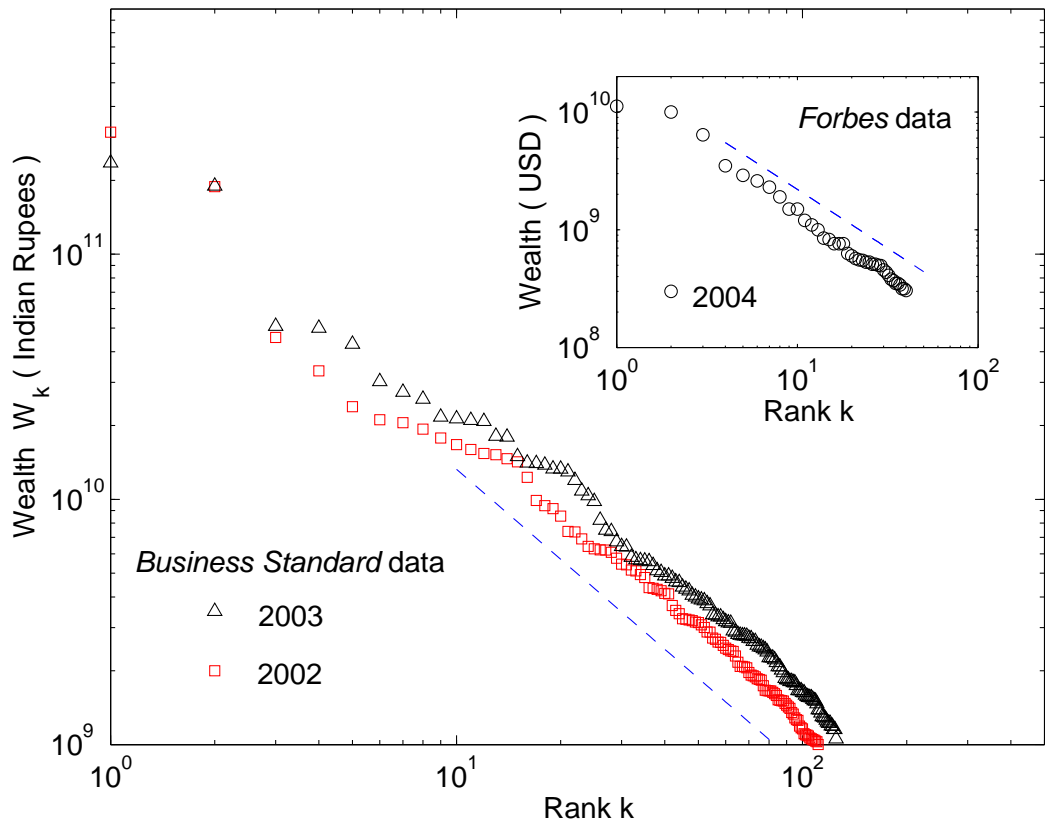}}
\caption{Rank ordered plots of the wealth of the richest Indians during
the period 2002-2004 on a double-logarithmic scale. 
The main figure shows the wealth of the $k$-th ranked
richest person (or household) 
against the rank $k$ (with rank 1 corresponding to the
wealthiest person) as per two surveys conducted by {\em Business Standard}
in Dec 31, 2002 (squares) and Aug 31, 2003 (triangles). The broken line
having a slope of $-1.23$ is shown for visual reference. The inset shows
the rank ordered plot of wealth based on data published by {\em Forbes}
in Dec 10, 2004, with the broken line having a slope of $-1.08$.
}
\label{fig1}
\end{figure}

\vspace{-0.3cm}
Fig.~\ref{fig1} shows the rank distribution of wealth from the lists of
richest Indians described in the previous section. Least square fit of the 
2002 data yields a slope of -1.24 while the 2003 data has a slope of -1.23,
which give Pareto exponents of 0.81 and 0.82, respectively. Note that,
due to the arrangement of the data, we could use only 111 points from the
2002 data, while all 125 data points could be used for the 2003 data.
The inset shows the 2004 data, which, upon least square fitting, gave a 
slope of -1.08, from which we obtained a Pareto exponent of 0.92.
Goodness of fit was quantitatively measured to be $R^2$ = 0.989 (2002),
0.984 (2003), and 0.988 (2004).

\begin{figure}
\epsfxsize=5.5in
\centerline{\epsffile{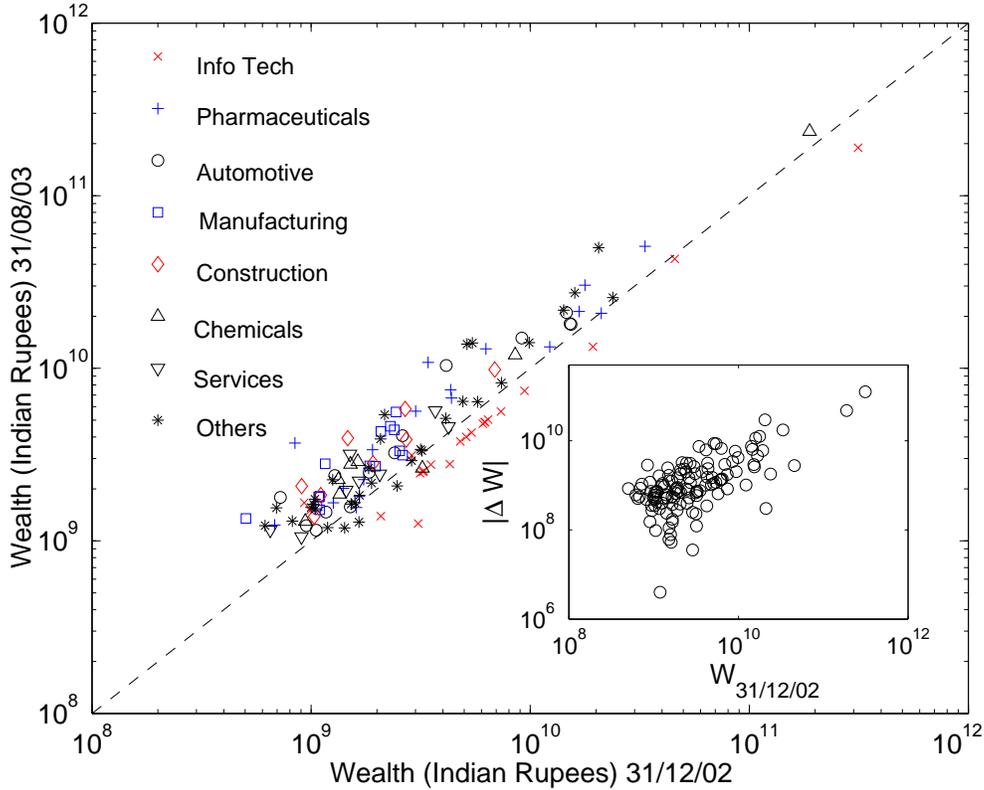}}
\caption{Wealth of the 125 richest Indians (as of Aug 31, 2003) compared
at two dates: Dec 31, 2002 and Aug 31, 2003, on a double-logarithmic scale. 
The data points are coded
according to the industry sector on which the wealth is based. The broken 
line corresponds to unchanged wealth over the period under study. 
The inset shows the absolute magnitude of change in wealth during this
period as a function of wealth at the start of the period.
}
\label{fig2}
\end{figure}

\vspace{-0.3cm}
Fig.~\ref{fig2} shows the correlation of net worth of agents over an interval of
6 months between Dec 31, 2002 and Aug 31, 2003. The points all fall in
a narrow band, implying that there is
no significant change in the wealth during this period. 
However, as all the fortunes being studied here are
based on stock holdings, movement in share values affect the net worth
of individuals (and households) in the list.
Fortunes based on information technology stocks show an uniform (although small)
decline over the period studied, whereas those based on
pharmaceuticals
stocks show, in general, an increase.

\vspace{-0.3cm}
The inset of Fig. 2 shows absolute changes in wealth over the period of 8
months as a function of the wealth at the beginning of the period.
The data points are all clustered close together, and the linear
correlation
coefficient in a log-log scale is 0.95, indicating that the wealth
lost or gained by agents is proportional to their overall wealth.
This is a characteristic of a multiplicative stochastic process,
where the changes in the value of a variable are proportional to the
value, rather than an additive process, where the changes are independent
of the value (e.g., random walks).
This lends support to the assumptions of asset exchange models for wealth
distribution \cite{Isp98,Dra00,Sin03,Cha04,Sla04,Sin05}, 
according to which, the amount lost or gained by agents through each
trading interaction
is a random fraction of their wealth at a given instant.

\begin{figure}
\epsfxsize=5.5in
\centerline{\epsffile{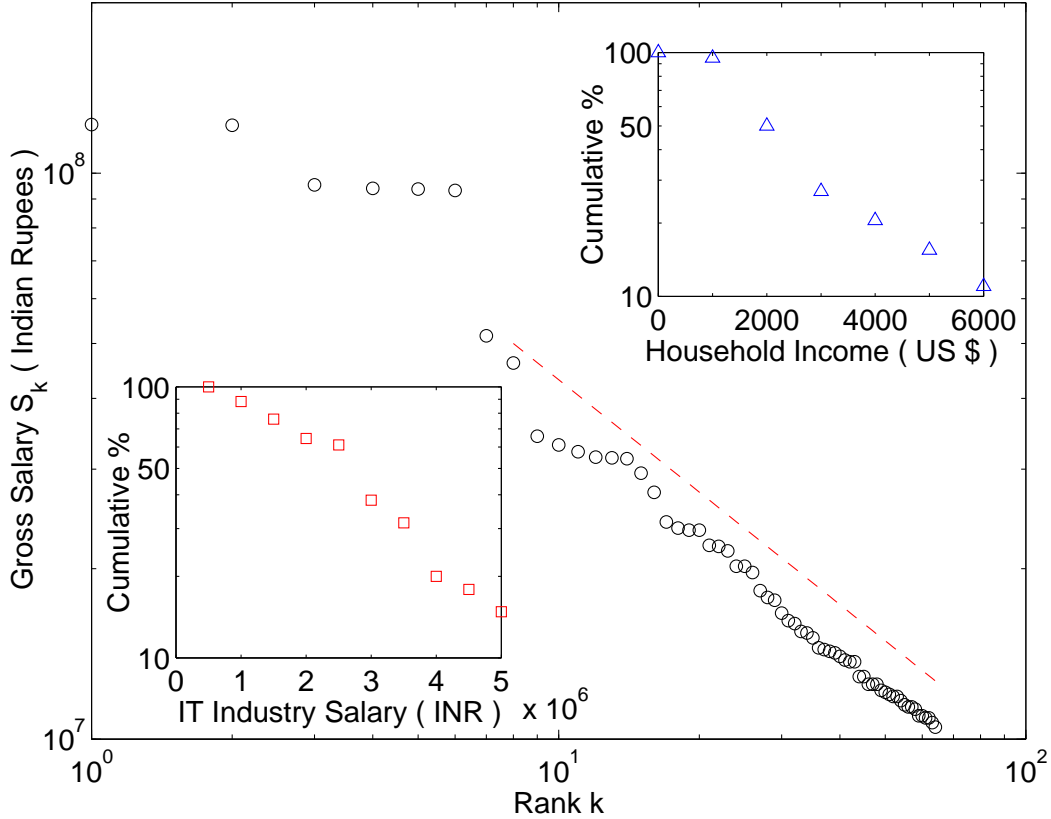}}
\caption{The rank ordered plot of the gross salary (in Indian Rupees) 
of the $k$-th ranked highest
paid executive against the rank $k$ on a double-logarithmic scale. The
broken line of slope $-0.66$ is shown for visual reference. 
The upper inset shows, on a semi-logarithmic scale, 
the cumulative percentage of Indian households at income
level $I$ (i.e., the percentage with household income greater than $I$)
plotted against $I$ (in US Dollars; 1 US Dollar $\simeq$ 37 Indian Rupees
during this period), 
for the lower-end of the income distribution. The lower
inset shows, on a semi-logarithmic scale, the percentage of
individuals in the Information Technology industry with 10 years or 
more experience, 
having a salary $S$ or more (in Indian Rupees).
}
\label{fig3}
\end{figure}

\vspace{-0.3cm}
Fig. \ref{fig3} shows the labor income (i.e., salaries) rank-order 
distribution for the highest paid company executives in India. 
Least square fitting of the data (rejecting the top 12.5 \% of the data points)
in a double-logarithmic scale gives
a slope of -0.66, which indicates a Pareto exponent $\alpha \simeq 1.51$
for the higher-end tail of labor income distribution in India. Note that,
this is almost identical to what Pareto had announced to be the value
of the exponent based on his study of late-19th century European
economies.
To compare the high-income end of the distribution
with the income of the rest of the population, the two insets show 
the cumulative income
distribution for lower-income Indian households (upper inset)
with data obtained from a 1997 survey
available online \cite{poverty}, and the cumulative salary distribution
for individuals (lower inset), with experience
of 10 years or more, working in the Information Technology industry,
where the data is from a 2002 survey by the IT industry magazine
{\em DataQuest} \cite{DataQ}. The data, although of low resolution,
is suggestive of a log-normal distribution in the low- to middle-income
range.

\vspace{-0.3cm}
Comparison between the overall income of poorer households and the salaries
(labor income) of middle- to high-income individuals is valid, because the
former comprises
almost entirely of wages, and not any earnings from financial or
other assets \cite{Sam01}.
It has been suggested that it is this difference in the composition of the
income between the low-income (comprising solely labor income) and 
high-income (dominated by capital investment gains) sections of the 
distribution that
is responsible for the exponential nature of the former and power-law
in the latter region \cite{Lev03}. 
However, we observe power-law even for the upper-end of the labor income
component of the high income individuals. This implies that the same process
may give rise to exponential behavior at the lower end of the distribution
while also being responsible for the power-law at the upper end, and models
for explaining the observed income distributions should satisfy this criterion.
\vspace{-0.7cm}
\section{Discussion}
\vspace{-0.7cm}
Based on the results reported above we conclude that
the Indian wealth distribution has a Pareto exponent between 0.81 and 0.92,
while the income distribution is log-normal with a power-law tail having 
a Pareto exponent close to 1.5, the value predicted by Pareto himself.
One should of course note that these values are not sacrosanct and that
there are several ways by which different values
of the Pareto exponent can be obtained for the same society.
For example, the Pareto exponent for the wealth distribution in UK has been
reported to have values as different as 1.9 \cite{Dra01}, 1.06 \cite{Lev98}
and 1.78 \cite{Coe04}. The data based on which these exponents were
obtained were of course for different years (1996,1997 and 2001, respectively);
however, that need not be the only reason for this striking discrepancy
among the values. For example, if the measured wealth consists solely
(or mostly) of
financial assets, in particular, stocks, as is likely for the wealths 
reported in the lists of
the richest published by {\em Business Standard} and {\em Fortune},
then the wealth inequality in a society is likely to be over-estimated
if middle-income households have a larger proportion of their wealth as
tangible assets (such as house or automobile) \cite{Wei97}. Thus, a study which
considers only financial assets is likely to come up with a Pareto 
exponent that differs substantially from another study that considers
the non-financial assets reported in data collected for the purpose
of calculating inheritance tax. 

\vspace{-0.3cm}
Another point worth considering is the relation
between Pareto exponent and Gini coefficient, the
most widely used measure of income inequality. According to the latter measure,
India is less unequal than USA, and even UK \cite{UN}. However, this is
not consistent with the measured values of Pareto exponent, if one
associates lower values of the exponent with increased inequality.
To resolve this issue, we note that if the distribution follows a 
power-law nature throughout, then a clear correspondence exists between
the two measures, e.g., a Pareto exponent of 1.5 implies a Gini coefficient
of 0.5 \cite{Per92}. However, observed distributions show a power-law
only over a very limited range, and hence the correspondence breaks down.
In fact, in this case, it has for long been a matter of debate whether a higher
value of Pareto exponent indicates increased or decreased income inequality
\cite{Per92}!


I thank Bikas K. Chakrabarti, Arnab Chatterjee and 
S. Subramanian for helpful suggestions.

\end{document}